\documentclass[%
superscriptaddress,
preprint,
amsmath,amssymb,
prc,
floatfix, ]%
{revtex4-1}

\usepackage{color}
\usepackage{CJK}
\usepackage{longtable}
\usepackage{graphicx}
\usepackage{dcolumn}
\usepackage{bm}
\usepackage[dvipdfm,bookmarks=true,colorlinks,%
            citecolor=blue,linkcolor=blue,hypertex, %
            breaklinks=true]{hyperref}

\begin{document}


\title{Systematic investigation of the high-$K$ isomers and the
       high-spin rotational bands in the neutron rich Nd and Sm isotopes
       by a particle-number conserving method}

\author{Zhen-Hua Zhang }
 \email{zhzhang@ncepu.edu.cn}
 \affiliation{Mathematics and Physics Department,
              North China Electric Power University, Beijing 102206, China}

\date{\today}

\begin{abstract}

The rotational properties of the neutron rich Nd and Sm isotopes
with mass number $A\approx150$ are systematically investigated using the cranked shell model
with pairing correlations treated by a particle-number conserving method,
in which the Pauli blocking effects are taken into account exactly.
The 2-quasiparticle states in even-even Nd and Sm isotopes with excitation energies lower
than 2.5~MeV are systematically calculated.
The available data can be well reproduced and some possible 2 and 4-quasiparticle isomers
are also suggested for future experiments.
The experimentally observed rotational frequency variations of moments of inertia
for the even-even and odd-$A$ nuclei are reproduced very well by the calculations.
The effects of high-order deformation $\varepsilon_6$ on the 2-quasiparticle
excitation energies and moments of inertia of the ground state bands in
even-even Nd and Sm isotopes are analyzed in detail.
By analyzing the occupation probability $n_\mu$ of each cranked
Nilsson orbitals near the Fermi surface and the contribution of
each major shell to the angular momentum alignments,
the alignment mechanism in these nuclei is understood clearly.

\end{abstract}


\maketitle

\section{Introduction}

For the neutron rich rare-earth nuclei with mass number $A\approx150$,
especially Nd ($Z=60$) and Sm ($Z=62$) isotopes, there are many novel phenomena,
e.g., nuclear quantum phase transition from spherical to deformed
shape~\cite{Iachello2004_PRL92-212501, Niksic2007_PRL99-092502},
octupole vibration~\cite{Phillips1986_PRL57-3257, Phillips1988_PLB212-402, Ibbotson1993_PRL71-1990},
$K$ isomers~\cite{Walker1999_Nature399-35}, etc.
From neutron number $N=92$, the nuclei are well-deformed and possess prolate ground state rotational bands.
In this mass region, there are several high-$K$ orbitals around the
proton and neutron Fermi surface, e.g., $\pi5/2^+[413]$, $\pi5/2^-[532]$, $\pi7/2^-[523]$,
$\nu5/2^-[523]$, $\nu5/2^+[642]$ and $\nu7/2^+[633]$.
Therefore, this may give rise to the formation of various high-$K$ multi-quasiparticle (qp) isomers,
which are particularly favorable for studying the blocking effects of the pairing correlations.

Due to the high statistics, the spontaneous fission of the actinide nuclei has been used to
populate the isomeric and high-spin states of neutron rich nuclei in $A\approx150$
mass region~\cite{Greenwood1987_PRC35-1965, Hamilton1995_PPNP35-635, Hamilton1997_PPNP38-273}.
Up to now, using the spontaneous fission of $^{252}$Cf~\cite{Zhu1995_JPG21-L57, Babu1996_PRC54-568,
Zhang1998_PRC57-2040, Gautherin1998_EPJA1-391, Hwang2008_PRC78-014309, Hwang2008_PRC78-017303,
Simpson2009_PRC80-024304, Urban2009_PRC80-037301, Hwang2010_PRC82-034308, Wang2014_PRC90-067306}
and in-flight fission of a $^{238}$U beam on a $^{9}$Be
target~\cite{Patel2014_PRL113-262502, Patel2016_PLB753-182, Ideguchi2016_PRC94-064322},
various high-$K$ isomers and high-spin rotational bands for the neutron rich
Nd and Sm isotopes, including both the even-even and the odd-$A$ nuclei, have been established.
Most recently, the lightest 4-qp high-$K$ isomer in this mass region has been observed
in $^{160}$Sm~\cite{ Patel2016_PLB753-182}.
These data can reveal detailed information on the single-particle structure,
shell structure, the high-$K$ isomerism, etc., thus providing a benchmark for various nuclear models.

Several nuclear models have been used to investigate the properties of these
neutron rich nuclei, including quasiparticle rotor model~\cite{Simpson2009_PRC80-024304, Urban2009_PRC80-037301},
a mean-field type Hartree-Fock-Bogoliubov theory with Gogny force D1S~\cite{Gautherin1998_EPJA1-391},
projected shell model~\cite{Yang2010_JPG37-085110, Yang2011_SciChinaPAM54-1}
and potential energy surface calculations~\cite{Patel2014_PRL113-262502, Patel2016_PLB753-182}.
However, most of these models focus on the even-even nuclei.
Only the projected shell model and the quasiparticle rotor model
were used to investigate the odd-$A$ nuclei $^{159}$Sm~\cite{Urban2009_PRC80-037301, Yang2010_JPG37-085110}.
Therefore, it is necessary to perform a systematic investigation including
both the even-even and the odd-$A$ Nd and Sm isotopes,
which can improve our understanding for these neutron rich nuclei.

In the present work, the cranked shell model (CSM) with
pairing correlations treated by a particle-number conserving (PNC)
method~\cite{Zeng1983_NPA405-1, Zeng1994_PRC50-1388} is used
to investigate systematically the neutron rich Nd and Sm isotopes with mass number
$A\approx150$, including both even-even and odd-$A$ nuclei.
In contrary to the conventional Bardeen-Cooper-Schrieffer or
Hartree-Fock-Bogoliubov approaches, the many-body Hamiltonian is solved directly
in a sufficiently large truncated Fock-space in the PNC method~\cite{Wu1989_PRC39-666}.
Therefore, the particle-number is conserved and the Pauli blocking effects are treated exactly.
The PNC-CSM has been employed successfully for describing various phenomena, e.g.,
the odd-even differences in moments of inertia (MOIs)~\cite{Zeng1994_PRC50-746},
identical bands~\cite{Liu2002_PRC66-024320, He2005_EPJA23-217},
nuclear pairing phase transition~\cite{Wu2011_PRC83-034323},
antimagnetic rotation~\cite{Zhang2013_PRC87-054314, Zhang2016_PRC94-034305},
rotational bands and high-$K$ isomers in the rare-earth~\cite{Liu2004_NPA735-77,
Zhang2009_NPA816-19, Zhang2009_PRC80-034313, Zhang2016_SciChinaPMA59-672012, Zhang2016_NPA949-22}
and actinide nuclei~\cite{He2009_NPA817-45,
Zhang2011_PRC83-011304R, Zhang2012_PRC85-014324, Zhang2013_PRC87-054308}, etc.
The PNC scheme has also been used both in relativistic
and non-relativistic mean field models~\cite{Meng2006_FPC1-38, Pillet2002_NPA697-141,Liang2015_PRC92-064325}
and the total-Routhian-surface method with the
Woods-Saxon potential~\cite{Fu2013_PRC87-044319, Fu2013_SCPMA56-1423}.
Recently, the PNC method based on the
cranking covariant density functional theory has been developed~\cite{Shi2018_PRC97-034317}.
Similar approaches to treat pairing correlations with exactly conserved
particle number can be found in Refs.~\cite{Richardson1964_NP52-221, Pan1998_PLB422-1,
Volya2001_PLB509-37, Jia2013_PRC88-044303, Jia2013_PRC88-064321, Chen2014_PRC89-014321}.

This paper is organized as follows.
A brief introduction to the PNC treatment of pairing correlations within
the CSM is presented in Sec.~\ref{Sec:PNC-CSM}.
The numerical details used in PNC calculations are given in Sec.~\ref{Sec:Numerical}.
In Sec.~\ref{Sec:Results}, the 2-qp energies and MOIs are calculated and compared with the data.
The 2-qp states in even-even Nd and Sm isotopes with excitation energies lower
than 2.5~MeV are systematically investigated.
The effects of high-order deformation $\varepsilon_6$ and alignment mechanism
in these nuclei are discussed in detail.
A brief summary is given in Sec.~\ref{Sec:Summary}.

\section{\label{Sec:PNC-CSM}PNC-CSM formalism}

The cranked shell model Hamiltonian of an axially symmetric
nucleus in the rotating frame can be written as
\begin{eqnarray}
 H_\mathrm{CSM}
 & = &
 H_0 + H_\mathrm{P}
 = H_{\rm Nil}-\omega J_x + H_\mathrm{P}
 \ ,
 \label{eq:H_CSM}
\end{eqnarray}
where $H_{\rm Nil}$ is the Nilsson Hamiltonian~\cite{Nilsson1969_NPA131-1},
$-\omega J_x$ is the Coriolis interaction with the cranking frequency $\omega$ about the
$x$ axis (perpendicular to the nuclear symmetry $z$ axis).
$H_{\rm P} = H_{\rm P}(0) + H_{\rm P}(2)$ is the pairing interaction,
\begin{eqnarray}
 H_{\rm P}(0)
 & = &
  -G_{0} \sum_{\xi\eta} a^\dag_{\xi} a^\dag_{\bar{\xi}}
                        a_{\bar{\eta}} a_{\eta}
  \ ,
 \\
 H_{\rm P}(2)
 & = &
  -G_{2} \sum_{\xi\eta} q_{2}(\xi)q_{2}(\eta)
                        a^\dag_{\xi} a^\dag_{\bar{\xi}}
                        a_{\bar{\eta}} a_{\eta}
  \ ,
\end{eqnarray}
where $\bar{\xi}$ ($\bar{\eta}$) labels the time-reversed state of a
Nilsson state $\xi$ ($\eta$),
$q_{2}(\xi) = \sqrt{{16\pi}/{5}}
\langle \xi |r^{2}Y_{20} | \xi \rangle$ is the diagonal element of
the stretched quadrupole operator,
and $G_0$ and $G_2$ are the effective strengths of monopole and
quadrupole pairing interaction, respectively.

Instead of the usual single-particle level truncation in conventional
shell-model calculations, a cranked many-particle configuration (CMPC)
truncation (Fock space truncation) is adopted, which is crucial
to make the PNC calculations for low-lying excited states both
workable and sufficiently accurate~\cite{Molique1997_PRC56-1795, Wu1989_PRC39-666}.
Usually a CMPC space with the dimension of 1000 should be enough for the calculations of rare-earth nuclei.
By diagonalizing the $H_\mathrm{CSM}$ in a sufficiently
large CMPC space, sufficiently accurate solutions for low-lying excited eigenstates of
$H_\mathrm{CSM}$ can be obtained, which can be written as
\begin{equation}
 |\Psi\rangle = \sum_{i} C_i \left| i \right\rangle
 \qquad (C_i \; \textrm{real}) \ ,
\end{equation}
where $| i \rangle$ is a CMPC (an eigenstate of the one-body operator $H_0$).

The angular momentum alignment for the state $| \Psi \rangle$ is
\begin{equation}
\langle \Psi | J_x | \Psi \rangle = \sum_i C_i^2 \langle i | J_x | i
\rangle + 2\sum_{i<j}C_i C_j \langle i | J_x | j \rangle \ ,
\end{equation}
and the kinematic MOI of state $| \Psi \rangle$ is
\begin{equation}
J^{(1)}=\frac{1}{\omega} \langle\Psi | J_x | \Psi \rangle \ .
\end{equation}
Because $J_x$ is a one-body operator, the matrix element $\langle i | J_x | j \rangle$
($i\neq j$) may not vanish only when
$|i\rangle$ and $|j\rangle$ differ by
one particle occupation~\cite{Zeng1994_PRC50-1388}.
After a certain permutation of creation operators,
$|i\rangle$ and $|j\rangle$ can be recast into
\begin{equation}
 |i\rangle=(-1)^{M_{i\mu}}|\mu\cdots \rangle \ , \qquad
|j\rangle=(-1)^{M_{j\nu}}|\nu\cdots \rangle \ ,
\end{equation}
where $\mu$ and $\nu$ denotes two different single-particle states,
and $(-1)^{M_{i\mu}}=\pm1$, $(-1)^{M_{j\nu}}=\pm1$ according to
whether the permutation is even or odd.
Therefore, the angular momentum alignment of
$|\Psi\rangle$ can be expressed as
\begin{equation}
 \langle \Psi | J_x | \Psi \rangle = \sum_{\mu} j_x(\mu) + \sum_{\mu<\nu} j_x(\mu\nu)
 \ .
 \label{eq:jx}
\end{equation}
where the diagonal contribution $j_x(\mu)$ and the
off-diagonal (interference) contribution $j_x(\mu\nu)$ can be written as
\begin{eqnarray}
j_x(\mu)&=&\langle\mu|j_{x}|\mu\rangle n_{\mu} \ ,
\\
j_x(\mu\nu)&=&2\langle\mu|j_{x}|\nu\rangle\sum_{i<j}(-1)^{M_{i\mu}+M_{j\nu}}C_{i}C_{j}
  \quad  (\mu\neq\nu) \ ,
\end{eqnarray}
and
\begin{equation}
n_{\mu}=\sum_{i}|C_{i}|^{2}P_{i\mu} \ ,
\end{equation}
is the occupation probability of the cranked orbital $|\mu\rangle$,
$P_{i\mu}=1$ if $|\mu\rangle$ is occupied in $|i\rangle$, and
$P_{i\mu}=0$ otherwise.

\section{\label{Sec:Numerical}Numerical details}

\begin{table*}[h]
  \centering
  \caption{Deformation parameters ($\varepsilon_2$, $\varepsilon_4$ and $\varepsilon_6$)
  of Nd and Sm isotopes used in present PNC-CSM calculations,
  which are taken from Ref.~\cite{Moeller1995_ADNDT59-185}.}\label{tab:def}
  \begin{tabular*}{1.0\textwidth}{@{\extracolsep{\fill}}ccccccccccc}
     \hline
   ~              & $^{152}$Nd & $^{153}$Nd & $^{154}$Nd & $^{155}$Nd & $^{156}$Nd & $^{157}$Nd & $^{158}$Nd & $^{159}$Nd & $^{160}$Nd \\
   \hline
  $\varepsilon_2$ & 0.242      & 0.250      & 0.250      & 0.258      & 0.258      & 0.258      & 0.258      & 0.258      & 0.267      \\
  $\varepsilon_4$ &-0.080      &-0.073      &-0.067      &-0.060      &-0.053      &-0.047      &-0.040      &-0.033      &-0.027      \\
  $\varepsilon_6$ & 0.026      & 0.031      & 0.034      & 0.037      & 0.038      & 0.040      & 0.040      & 0.042      & 0.043      \\
  \hline
   ~              & $^{154}$Sm & $^{155}$Sm & $^{156}$Sm & $^{157}$Sm & $^{158}$Sm & $^{159}$Sm & $^{160}$Sm & $^{161}$Sm & $^{162}$Sm \\
   \hline
  $\varepsilon_2$ & 0.250      & 0.250      & 0.258      & 0.258      & 0.258      & 0.267      & 0.267      & 0.275      & 0.275      \\
  $\varepsilon_4$ &-0.067      &-0.060      &-0.053      &-0.047      &-0.040      &-0.033      &-0.027      &-0.013      &-0.007      \\
  $\varepsilon_6$ & 0.030      & 0.032      & 0.038      & 0.038      & 0.040      & 0.044      & 0.044      & 0.045      & 0.046      \\
   \hline
   \end{tabular*}
\end{table*}

In this work, the deformation parameters ($\varepsilon_2$, $\varepsilon_4$ and $\varepsilon_6$)
of Nd and Sm isotopes used in PNC-CSM calculations are taken from Ref.~\cite{Moeller1995_ADNDT59-185},
which are shown at Table~\ref{tab:def}.
The Nilsson parameters ($\kappa$ and $\mu$) are taken as
the traditional values~\cite{Nilsson1969_NPA131-1}.
The experimental data show that the ground state of $N=93$
isotones (e.g., $^{153}$Nd and $^{155}$Sm) is $\nu3/2^-[521]$~\cite{Hwang1997_IJMPE6-331, Reich2005_NDS104-1}.
However, the calculated ground state using the traditional
Nilsson parameters is $\nu5/2^+[642]$~\cite{Nilsson1969_NPA131-1}.
Therefore, to reproduce the experimental level sequence,
the neutron orbital $\nu5/2^+[642]$ is shifted upwards
slightly by $0.07\hbar\omega_0$ for all these nuclei.

\begin{figure}[h]
\includegraphics[width=0.7\columnwidth]{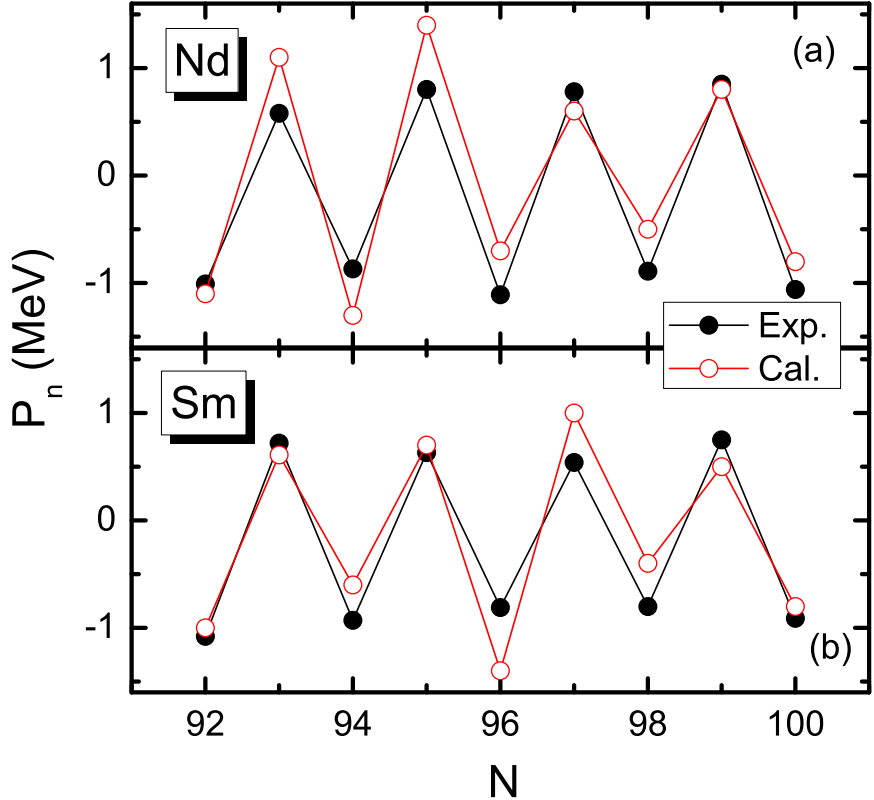}
\caption{\label{fig0:oe}(Color online)
The comparison between the experimental (black solid circles) and calculated (red open circles)
neutron odd-even difference $P_n$ for Nd (upper panel) and Sm (lower panel) isotopes.
The experimental binding energies are taken from Ref.~\cite{Wang2012_ChinPhysC36-1603}.
}
\end{figure}

The effective pairing strengths for each nuclei, in principle, can be determined by
the experimental odd-even differences in nuclear binding energies~\cite{Wang2012_ChinPhysC36-1603},
\begin{eqnarray}
P_{\rm p} &=& \frac{1}{2}\left[ B(Z+1, N)+B(Z-1, N)\right] - B(Z, N) \nonumber\\
          &=&  E_{\rm g}(Z, N) - \frac{1}{2}\left[  E_{\rm g}(Z+1, N) + E_{\rm g}(Z-1, N)\right] \nonumber \ , \\
P_{\rm n} &=& \frac{1}{2}\left[ B(Z, N+1)+B(Z, N-1)\right] - B(Z, N) \nonumber\\
          &=&  E_{\rm g}(Z, N) - \frac{1}{2}\left[  E_{\rm g}(Z, N+1) + E_{\rm g}(Z, N-1)\right] \ ,
          \label{eq:oed}
\end{eqnarray}
where $E_{\rm g}$ is the ground state energy of the nucleus,
and are connected with the dimension of the truncated CMPC space.
In this work, the CMPC space is constructed in the proton $N=4, 5$ major shells
and the neutron $N=5, 6$ major shells, respectively.
The CMPC truncation energies are about 0.85$\hbar\omega_0$ for protons and 0.80$\hbar\omega_0$ for neutrons, respectively.
The dimensions of the CMPC space are about 1000 for both protons and neutrons in the present calculation.
For all Nd and Sm isotopes, the corresponding effective monopole and quadrupole pairing strengths are chosen as
$G_{\rm 0p}=0.25$~MeV and $G_{\rm 2p}=0.01$~MeVfm$^{-4}$ for protons,
$G_{\rm 0n}=0.30$~MeV and $G_{\rm 2n}=0.02$~MeVfm$^{-4}$ for neutrons.
Figure~\ref{fig0:oe} shows the comparison between
experimental (black solid circles) and calculated (red open circles) neutron
odd-even difference $P_n$ for Nd (upper panel) and Sm (lower panel) isotopes.
It can be seen that the data can be well reproduced.
In principal, the pairing strengths should be different for each nucleus.
Note that for some neutron rich Nd and Sm nuclei,
the experimental binding energy is not accurate~\cite{Wang2012_ChinPhysC36-1603}.
Therefore, in the present work the pairing strengths for
all nuclei are chosen as the same value to get a global fit.
Previous investigations have shown that
after the quadrupole pairing being included,
the description of experimental band-head energies
and the level crossing frequencies can be improved~\cite{Diebel1984_NPA419-221}.
As for the quadrupole pairing, the strength is determined by the bandhead MOIs in the present work.
The quarople pairing is also included in the projected shell model
when investigating the Nd and Sm isotopes, in which BCS method is used to
treat the pairing correlations~\cite{Yang2010_JPG37-085110}.
In the projected shell model, the quadrupole pairing strength
is chosen to be proportional to the monopole pairing strength with
proportionality constant 0.18~\cite{Yang2010_JPG37-085110}.
As for this point, this proportionality is much smaller
in the present work (less than 0.1).
However, the PNC method is different from the BCS method.
Therefore, the effective pairing strength should be different.

The stability of the calculations against the change of the dimension of
the CMPC space has been investigated in
Refs.~\cite{Molique1997_PRC56-1795,Zeng1994_PRC50-1388, Zhang2012_PRC85-014324}.
In present calculations, almost all the CMPCs with weight $>0.1\%$ in the many-body wave functions
are taken into account, so the solutions to the low-lying excited states are accurate enough.
A larger CMPC space with renormalized pairing strengths gives essentially the same results.

\section{\label{Sec:Results}Results and discussion}

\subsection{Cranked Nilsson levels}

\begin{figure}[h]
\includegraphics[width=0.9\columnwidth]{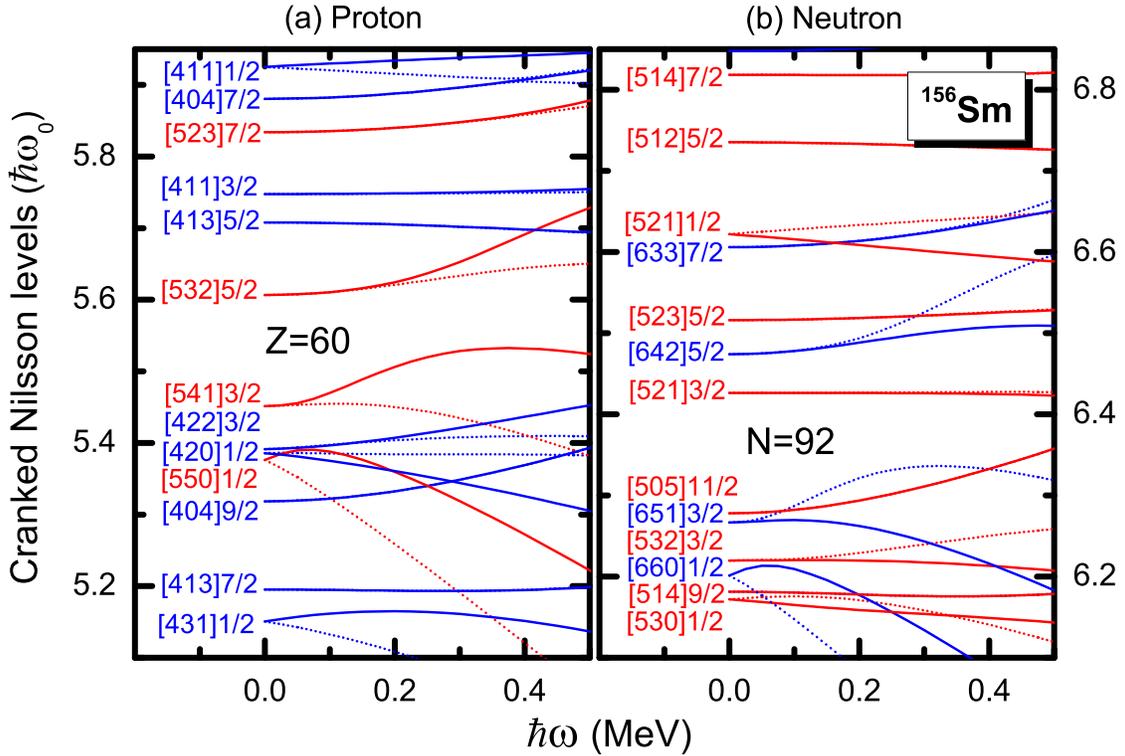}
\caption{\label{fig1:Nil}(Color online)
The cranked Nilsson levels near the Fermi surface of $^{156}$Sm
for (a) protons and (b) neutrons.
The positive (negative) parity levels are denoted by blue (red) lines.
The signature $\alpha=+1/2$ ($\alpha=-1/2$) levels
are denoted by solid (dotted) lines.
The Nilsson parameters ($\kappa$ and $\mu$) are taken as
the traditional values~\cite{Nilsson1969_NPA131-1}.
The deformation parameters $\varepsilon_2 = 0.258$, $\varepsilon_4 = -0.053$
and $\varepsilon_6 = 0.038$ are taken from Ref.~\cite{Moeller1995_ADNDT59-185}.
In addition, the neutron orbital $\nu5/2^+[642]$ is shifted upwards slightly by $0.07\hbar\omega_0$
to reproduce the experimental level sequence.}
\end{figure}

As an example of Nd and Sm isotopes around $A\approx150$ mass region,
the calculated cranked Nilsson levels near the Fermi surface of $^{156}$Sm are shown in Fig.~\ref{fig1:Nil}.
The positive (negative) parity levels are denoted by blue (red) lines.
The signature $\alpha=+1/2$ ($\alpha=-1/2$) levels
are denoted by solid (dotted) lines.
It can be seen from Fig.~\ref{fig1:Nil} that, there are several high-$K$ orbitals around the
proton and neutron Fermi surface, e.g., $\pi5/2^+[413]$, $\pi5/2^-[532]$, $\pi7/2^-[523]$,
$\nu5/2^-[523]$, $\nu5/2^+[642]$ and $\nu7/2^+[633]$.
Therefore, this may lead to the formation of various high-$K$
multi-qp isomers in Nd and Sm isotopes around $A\approx150$ mass region.
It also can be seen that there are two sub-shells at proton number $Z=60$ and neutron number $N=92$, respectively.
So for Nd ($Z=60$) isotopes, the excitation energies of the proton 2-qp states should be a little higher.
The experimentally favored 2-qp states should base on neutron configurations.
Indeed, no proton 2-qp state in neutron rich Nd isotopes has been observed experimentally up to now.
The energy gap at $Z=62$ is much smaller than that of $Z=60$.
So for Sm ($Z=62$) isotopes, the proton 2-qp states should exist.

\subsection{2-qp excitation energies in even-even Nd and Sm isotopes}

A series of 2-qp isomers have been observed experimentally in even-even
Nd and Sm isotopes at $A\approx150$ mass region~\cite{Simpson2009_PRC80-024304,
Patel2014_PRL113-262502, Ideguchi2016_PRC94-064322, Wang2014_PRC90-067306, Patel2016_PLB753-182},
which provide detailed information for these neutron rich nuclei.
It should be noted that in this mass region, the high-order deformation $\varepsilon_6$ is
remarkable~\cite{Moeller1995_ADNDT59-185}, and have a measurable effect on
the structure of these nuclei, e.g.,
the inclusion of $\varepsilon_6$ will alter the
the 2-qp excitation energy about 250~keV~\cite{Patel2014_PRL113-262502}.
Systematically calculated 2-qp states in even-even Nd isotopes with excitation energies lower
than 2.5~MeV are shown in Table~\ref{tab:2qpNd}.
$E_{\rm cal}$ and $E_{\rm cal}^*$ denote the calculated results with
and without $\varepsilon_6$ deformation, respectively.
In addition, the energy differences $|\Delta E|= |E_{\rm cal}-E_{\rm cal}^*|$
are also shown in the last column.
It can be seen that, the data are reproduced quite well by the PNC-CSM calculations no matter
whether the $\varepsilon_6$ deformation is considered or not.
This indicates that the adopted single-particle level scheme is suitable for the PNC-CSM calculations.
The energy differences $|\Delta E|$ for these four observed 2-qp states are all less than 100~keV.
It seems that the $\varepsilon_6$ deformation has small effects
on the excitation energies of these 2-qp states.
However, if one see through Table~\ref{tab:2qpNd}, the effects
of $\varepsilon_6$ deformation are prominent in some 2-qp states.
For example, the excitation energy of  $3^+$ state with configuration
$\nu^2 1/2^-[521]\otimes 5/2^-[512]$ in $^{160}$Nd is lowered by 439~keV after the
$\varepsilon_6$ deformation being neglected.
This is because after the $\varepsilon_6$ deformation is switched off,
the sequence of the single-particle levels is changed.
The root-mean-square deviation between $E_{\rm cal}$ and $E_{\rm cal}^*$ is about 130~keV.
Therefore, the $\varepsilon_6$ deformation still has remarkable effects on the excitation energies of
the 2-qp states.
Due to the large shell gap at $Z=60$, the energies of the proton 2-qp states
in Nd isotopes are all quite high.
However, with increasing neutron number, the energy of the lowest proton 2-qp state
with $K^\pi=4^+$ in each Nd isotopes decreases from more than 2.0~MeV to about 1.6~MeV,
which may be observed in future experiments.
The lowering of the excitation energy of $K^\pi=4^+$ state is caused by the decreasing of
the $Z=60$ shell gap with increasing neutron number.
Since the proton-neutron residual interaction is neglected in the PNC-CSM calculations,
the excitation energies of the 4-qp states with two quasi-protons and two quasi-neutrons
can be simply obtained by summing the energies of the corresponding 2-qp states.
Especially for $^{160}$Nd, the excitation energy of the 4-qp state
$K^\pi=8^-$ ($\pi^2 3/2^-[541]5/2^-[532]\otimes \nu^2 1/2^-[521]7/2^+[633]$)
is only about 2899~keV from PNC-CSM calculation.
Therefore, I hope this state can be observed by future experiments.

\begin{longtable}[h]{@{\extracolsep{\fill}}lcccccr}
  \caption{2-qp states in even-even Nd isotopes calculated by PNC-CSM with ($E_{\rm cal}$) and
  without ($E_{\rm cal}^*$) $\varepsilon_6$ deformation, where $|\Delta E|= |E_{\rm cal}-E_{\rm cal}^*|$.
  The data are taken from Refs.~\cite{Simpson2009_PRC80-024304, Ideguchi2016_PRC94-064322}.}
  \label{tab:2qpNd}\\[0mm]
   \hline
    Nucleus    & $K^\pi$ & Configuration & $E_{\rm exp}$ (keV) & $E_{\rm cal}$ (keV)& $E_{\rm cal}^*$ (keV) & $|\Delta E|$\\
    \hline
    $^{152}$Nd & $4^+$   & $\pi^2\frac{3}{2}^-[541]\otimes\frac{5}{2}^-[532]$ &      & 2038 & 2177 & 139 \\
    $^{152}$Nd & $4^-$   & $\pi^2\frac{3}{2}^+[422]\otimes\frac{5}{2}^-[532]$ &      & 2107 & 2148 & 41  \\
    $^{152}$Nd & $3^-$   & $\pi^2\frac{1}{2}^+[420]\otimes\frac{5}{2}^-[532]$ &      & 2325 & 2398 & 73  \\
    \hline

    $^{154}$Nd & $4^-$   & $\nu^2\frac{3}{2}^-[521]\otimes\frac{5}{2}^+[642]$ & 1298~\cite{Simpson2009_PRC80-024304} & 1263 & 1307 & 44  \\
    $^{154}$Nd & $4^+$   & $\nu^2\frac{3}{2}^-[521]\otimes\frac{5}{2}^-[523]$ &      & 1575 & 1550 & 25  \\
    $^{154}$Nd & $5^-$   & $\nu^2\frac{5}{2}^-[523]\otimes\frac{5}{2}^+[642]$ &      & 1921 & 2026 & 105 \\
    $^{154}$Nd & $5^-$   & $\nu^2\frac{3}{2}^-[521]\otimes\frac{7}{2}^+[633]$ &      & 2333 & 2396 & 63  \\
    $^{154}$Nd & $2^+$   & $\nu^2\frac{1}{2}^-[521]\otimes\frac{3}{2}^-[521]$ &      & 2400 & 2346 & 54  \\
    $^{154}$Nd & $4^+$   & $\pi^2\frac{3}{2}^-[541]\otimes\frac{5}{2}^-[532]$ &      & 1906 & 2090 & 184 \\
    $^{154}$Nd & $4^-$   & $\pi^2\frac{3}{2}^+[422]\otimes\frac{5}{2}^-[532]$ &      & 2085 & 2127 & 42  \\
    $^{154}$Nd & $3^-$   & $\pi^2\frac{1}{2}^+[420]\otimes\frac{5}{2}^-[532]$ &      & 2222 & 2310 & 88  \\
    \hline

    $^{156}$Nd & $5^-$   & $\nu^2\frac{5}{2}^-[523]\otimes\frac{5}{2}^+[642]$ & 1431~\cite{Simpson2009_PRC80-024304} & 1437 & 1351 & 86  \\
    $^{156}$Nd & $4^+$   & $\nu^2\frac{3}{2}^-[521]\otimes\frac{5}{2}^-[523]$ &      & 1501 & 1555 & 54  \\
    $^{156}$Nd & $6^+$   & $\nu^2\frac{5}{2}^+[642]\otimes\frac{7}{2}^+[633]$ &      & 2075 & 2152 & 77  \\
    $^{156}$Nd & $4^-$   & $\nu^2\frac{3}{2}^-[521]\otimes\frac{5}{2}^+[642]$ &      & 2086 & 1970 & 116 \\
    $^{156}$Nd & $5^-$   & $\nu^2\frac{3}{2}^-[521]\otimes\frac{7}{2}^+[633]$ &      & 2179 & 2341 & 162 \\
    $^{156}$Nd & $3^-$   & $\nu^2\frac{1}{2}^-[521]\otimes\frac{5}{2}^+[642]$ &      & 2204 & 2143 & 61  \\
    $^{156}$Nd & $6^-$   & $\nu^2\frac{5}{2}^-[523]\otimes\frac{7}{2}^+[633]$ &      & 2292 & 2295 & 3   \\
    $^{156}$Nd & $2^+$   & $\nu^2\frac{1}{2}^-[521]\otimes\frac{3}{2}^-[521]$ &      & 2378 & 2285 & 93  \\
    $^{156}$Nd & $4^+$   & $\pi^2\frac{3}{2}^-[541]\otimes\frac{5}{2}^-[532]$ &      & 1863 & 1983 & 120 \\
    $^{156}$Nd & $4^-$   & $\pi^2\frac{3}{2}^+[422]\otimes\frac{5}{2}^-[532]$ &      & 2130 & 2096 & 34  \\
    $^{156}$Nd & $3^-$   & $\pi^2\frac{1}{2}^+[420]\otimes\frac{5}{2}^-[532]$ &      & 2188 & 2224 & 36  \\
    \hline

    $^{158}$Nd & $6^-$   & $\nu^2\frac{5}{2}^-[523]\otimes\frac{7}{2}^+[633]$ & 1648~\cite{Ideguchi2016_PRC94-064322} & 1557 & 1595 & 38  \\
    $^{158}$Nd & $3^+$   & $\nu^2\frac{1}{2}^-[521]\otimes\frac{5}{2}^-[523]$ &      & 1745 & 1632 & 113 \\
    $^{158}$Nd & $6^+$   & $\nu^2\frac{5}{2}^+[642]\otimes\frac{7}{2}^+[633]$ &      & 2104 & 2004 & 100 \\
    $^{158}$Nd & $5^-$   & $\nu^2\frac{3}{2}^-[521]\otimes\frac{7}{2}^+[633]$ &      & 2263 & 2275 & 12  \\
    $^{158}$Nd & $3^-$   & $\nu^2\frac{1}{2}^-[521]\otimes\frac{5}{2}^+[642]$ &      & 2283 & 2039 & 244 \\
    $^{158}$Nd & $2^+$   & $\nu^2\frac{1}{2}^-[521]\otimes\frac{3}{2}^-[521]$ &      & 2442 & 2309 & 133 \\
    $^{158}$Nd & $4^+$   & $\pi^2\frac{3}{2}^-[541]\otimes\frac{5}{2}^-[532]$ &      & 1699 & 1862 & 163 \\
    $^{158}$Nd & $3^-$   & $\pi^2\frac{1}{2}^+[420]\otimes\frac{5}{2}^-[532]$ &      & 2077 & 2158 & 81  \\
    $^{158}$Nd & $4^-$   & $\pi^2\frac{3}{2}^+[422]\otimes\frac{5}{2}^-[532]$ &      & 2097 & 2105 & 8   \\
    $^{158}$Nd & $4^-$   & $\pi^2\frac{5}{2}^+[413]\otimes\frac{3}{2}^-[541]$ &      & 2381 & 2173 & 208 \\
    \hline

    $^{160}$Nd & $4^-$   & $\nu^2\frac{1}{2}^-[521]\otimes\frac{7}{2}^+[633]$ & 1108~\cite{Ideguchi2016_PRC94-064322} & 1243 & 1258 & 15  \\
    $^{160}$Nd & $3^+$   & $\nu^2\frac{1}{2}^-[521]\otimes\frac{5}{2}^-[523]$ &      & 1845 & 1991 & 146 \\
    $^{160}$Nd & $6^-$   & $\nu^2\frac{5}{2}^-[523]\otimes\frac{7}{2}^+[633]$ &      & 2117 & 2037 & 80  \\
    $^{160}$Nd & $6^-$   & $\nu^2\frac{5}{2}^-[512]\otimes\frac{7}{2}^+[633]$ &      & 2215 & 1941 & 274 \\
    $^{160}$Nd & $3^-$   & $\nu^2\frac{1}{2}^-[521]\otimes\frac{5}{2}^+[642]$ &      & 2412 & 2435 & 23  \\
    $^{160}$Nd & $3^+$   & $\nu^2\frac{1}{2}^-[521]\otimes\frac{5}{2}^-[512]$ &      & 2451 & 2012 & 439 \\
    $^{160}$Nd & $4^+$   & $\pi^2\frac{3}{2}^-[541]\otimes\frac{5}{2}^-[532]$ &      & 1656 & 1797 & 141 \\
    $^{160}$Nd & $3^-$   & $\pi^2\frac{1}{2}^+[420]\otimes\frac{5}{2}^-[532]$ &      & 2033 & 2111 & 78  \\
    $^{160}$Nd & $4^-$   & $\pi^2\frac{3}{2}^+[422]\otimes\frac{5}{2}^-[532]$ &      & 2127 & 2126 & 1   \\
    $^{160}$Nd & $7^-$   & $\pi^2\frac{9}{2}^+[404]\otimes\frac{5}{2}^-[532]$ &      & 2336 & 2516 & 180 \\
   \hline
\end{longtable}

In Table~\ref{tab:2qpSm}, similar results are shown for Sm isotopes.
Different from Nd isotopes, the shell gap at $Z=62$ is much smaller, so the proton 2-qp states should exist.
Indeed, 2-qp states with $K^\pi=5^-$ based on proton
configuration $\pi^2 5/2^+[413]\otimes5/2^-[532]$ have been observed in
$^{158}$Sm~\cite{Wang2014_PRC90-067306} and $^{160}$Sm~\cite{Patel2016_PLB753-182}.
The available data are also reproduced quite well by the PNC-CSM calculations,
especially after the $\varepsilon_6$ deformation being considered.
From Table~\ref{tab:2qpSm} it can be seen that, the effects of $\varepsilon_6$ deformation on
Sm isotopes are more prominent than Nd isotopes.
The root-mean-square deviation between $E_{\rm cal}$ and $E_{\rm cal}^*$ is about 260~keV,
which is consistent with the potential energy surface calculations in Ref.~\cite{Patel2014_PRL113-262502}.
In Ref.~\cite{Simpson2009_PRC80-024304}, the calculated lowest 2-qp state in $^{156}$Sm
using the quasiparticle rotor model is $K^\pi=4^-$.
However, in their calculation, the $K^\pi=5^-$ state is yrast with increasing spin.
Therefore, they assigned the 1398keV state in $^{156}$Sm as $K^\pi=5^-$.
In the PNC-CSM calculations, the excitation energy of $K^\pi=5^-$ is much higher than 1398~keV,
whereas the calculated $K^\pi=4^-$ is very close to the data.
In addition, the excitation energies of the two $K^\pi=5^-$ states in $^{158}$Sm are quite close to each other,
so their configurations need further investigation.
These will be discussed later.
Recently, one 4-qp isomer with excitation energy 2757~keV has been observed in $^{160}$Sm, which is assigned as
$K^\pi=11^-$ ($\pi^2 5/2^+[413]5/2^-[532]\otimes \nu^2 5/2^-[523]7/2^+[633]$).
It can be seen that the proton 2-qp state $\pi^2 5/2^+[413]5/2^-[532]$ and neutron
2-qp state $\nu^2 5/2^-[523]7/2^+[633]$ are all the lowest-lying 2-qp excitations in $^{160}$Sm.
The calculated excitation energy for this 4-qp state is 2918~keV, which is very close to the data.
Due to the low excitation energy of the proton 2-qp states in Sm isotopes, possible 4-qp states
with the lowest 2-quasi-proton and 2-quasi-neutron configurations may exist.

\begin{longtable}[h]{@{\extracolsep{\fill}}lcccccr}
  \caption{Similar as Table~\ref{tab:2qpNd}, but for even-even Sm isotopes.
  The data are taken from Refs.~\cite{Simpson2009_PRC80-024304, Wang2014_PRC90-067306, Patel2016_PLB753-182}.
  The 1398~keV state in $^{156}$Sm was assigned as $K^\pi=5^-$ in Ref.~\cite{Simpson2009_PRC80-024304}.}
  \label{tab:2qpSm}\\
   \hline
    Nucleus    & $K^\pi$ & Configuration & $E_{\rm exp}$ (keV) & $E_{\rm cal}$ (keV)& $E_{\rm cal}^*$ (keV) & $|\Delta E|$\\
    \hline
    $^{154}$Sm & $5^-$   & $\pi^2\frac{5}{2}^+[413]\otimes\frac{5}{2}^-[532]$ &        & 1400 & 1182 & 218 \\
    $^{154}$Sm & $4^-$   & $\pi^2\frac{3}{2}^+[411]\otimes\frac{5}{2}^-[532]$ &        & 1664 & 1455 & 209 \\
    $^{154}$Sm & $4^+$   & $\pi^2\frac{3}{2}^+[411]\otimes\frac{5}{2}^+[413]$ &        & 2331 & 1834 & 497 \\
    $^{154}$Sm & $6^+$   & $\pi^2\frac{5}{2}^-[532]\otimes\frac{7}{2}^-[523]$ &        & 2372 & 2337 & 35  \\
    $^{154}$Sm & $6^-$   & $\pi^2\frac{7}{2}^+[404]\otimes\frac{5}{2}^-[532]$ &        & 2475 & 2369 & 106 \\
    \hline

    $^{156}$Sm & $4^-$   & $\nu^2\frac{3}{2}^-[521]\otimes\frac{5}{2}^+[642]$ & (1398)~\cite{Simpson2009_PRC80-024304} & 1386 & 1354 & 32  \\
    $^{156}$Sm & $4^+$   & $\nu^2\frac{3}{2}^-[521]\otimes\frac{5}{2}^-[523]$ &        & 1642 & 1527 & 115 \\
    $^{156}$Sm & $5^-$   & $\nu^2\frac{5}{2}^-[523]\otimes\frac{5}{2}^+[642]$ &        & 1981 & 2040 & 59  \\
    $^{156}$Sm & $5^-$   & $\nu^2\frac{3}{2}^-[521]\otimes\frac{7}{2}^+[633]$ &        & 2335 & 2326 & 9   \\
    $^{156}$Sm & $2^+$   & $\nu^2\frac{1}{2}^-[521]\otimes\frac{3}{2}^-[521]$ &        & 2467 & 2319 & 148 \\
    $^{156}$Sm & $5^-$   & $\pi^2\frac{5}{2}^+[413]\otimes\frac{5}{2}^-[532]$ &        & 1444 & 1200 & 244 \\
    $^{156}$Sm & $4^-$   & $\pi^2\frac{3}{2}^+[411]\otimes\frac{5}{2}^-[532]$ &        & 1699 & 1481 & 218 \\
    $^{156}$Sm & $6^+$   & $\pi^2\frac{5}{2}^-[532]\otimes\frac{7}{2}^-[523]$ &        & 2344 & 2308 & 36  \\
    $^{156}$Sm & $4^+$   & $\pi^2\frac{3}{2}^+[411]\otimes\frac{5}{2}^+[413]$ &        & 2408 & 1855 & 553 \\
    \hline

    $^{158}$Sm & $5^-$   & $\nu^2\frac{5}{2}^-[523]\otimes\frac{5}{2}^+[642]$ & 1279~\cite{Simpson2009_PRC80-024304}   & 1394 & 1305 & 89  \\
    $^{158}$Sm & $4^+$   & $\nu^2\frac{3}{2}^-[521]\otimes\frac{5}{2}^-[523]$ &        & 1508 & 1571 & 63  \\
    $^{158}$Sm & $6^+$   & $\nu^2\frac{5}{2}^+[642]\otimes\frac{7}{2}^+[633]$ &        & 2015 & 2103 & 88  \\
    $^{158}$Sm & $4^-$   & $\nu^2\frac{3}{2}^-[521]\otimes\frac{5}{2}^+[642]$ &        & 2005 & 1887 & 118 \\
    $^{158}$Sm & $6^-$   & $\nu^2\frac{5}{2}^-[523]\otimes\frac{7}{2}^+[633]$ &        & 2142 & 2144 & 2   \\
    $^{158}$Sm & $5^-$   & $\nu^2\frac{3}{2}^-[521]\otimes\frac{7}{2}^+[633]$ &        & 2147 & 2344 & 197 \\
    $^{158}$Sm & $3^-$   & $\nu^2\frac{1}{2}^-[521]\otimes\frac{5}{2}^+[642]$ &        & 2205 & 2134 & 71  \\
    $^{158}$Sm & $3^+$   & $\nu^2\frac{1}{2}^-[521]\otimes\frac{5}{2}^-[523]$ &        & 2332 & 2180 & 152 \\
    $^{158}$Sm & $2^+$   & $\nu^2\frac{1}{2}^-[521]\otimes\frac{3}{2}^-[521]$ &        & 2347 & 2378 & 31  \\
    $^{158}$Sm & $5^-$   & $\pi^2\frac{5}{2}^+[413]\otimes\frac{5}{2}^-[532]$ & 1322~\cite{Wang2014_PRC90-067306}   & 1384 & 1197 & 187 \\
    $^{158}$Sm & $4^-$   & $\pi^2\frac{3}{2}^+[411]\otimes\frac{5}{2}^-[532]$ &        & 1665 & 1472 & 193 \\
    $^{158}$Sm & $6^+$   & $\pi^2\frac{5}{2}^-[532]\otimes\frac{7}{2}^-[523]$ &        & 2279 & 2272 & 7   \\
    $^{158}$Sm & $4^+$   & $\pi^2\frac{3}{2}^+[411]\otimes\frac{5}{2}^+[413]$ &        & 2329 & 1762 & 567 \\
    \hline

    $^{160}$Sm & $6^-$   & $\nu^2\frac{5}{2}^-[523]\otimes\frac{7}{2}^+[633]$ & 1468~\cite{Patel2016_PLB753-182}   & 1457 & 1730 & 273 \\
    $^{160}$Sm & $3^+$   & $\nu^2\frac{1}{2}^-[521]\otimes\frac{5}{2}^-[523]$ &        & 1697 & 1840 & 143 \\
    $^{160}$Sm & $6^+$   & $\nu^2\frac{5}{2}^+[642]\otimes\frac{7}{2}^+[633]$ &        & 2062 & 2150 & 88  \\
    $^{160}$Sm & $5^-$   & $\nu^2\frac{3}{2}^-[521]\otimes\frac{7}{2}^+[633]$ &        & 2192 & 2376 & 184 \\
    $^{160}$Sm & $3^-$   & $\nu^2\frac{1}{2}^-[521]\otimes\frac{5}{2}^+[642]$ &        & 2216 & 2220 & 4   \\
    $^{160}$Sm & $2^+$   & $\nu^2\frac{1}{2}^-[521]\otimes\frac{3}{2}^-[521]$ &        & 2474 & 2447 & 27  \\
    $^{160}$Sm & $5^-$   & $\pi^2\frac{5}{2}^+[413]\otimes\frac{5}{2}^-[532]$ & 1361~\cite{Patel2016_PLB753-182}   & 1461 & 1159 & 302 \\
    $^{160}$Sm & $4^-$   & $\pi^2\frac{3}{2}^+[411]\otimes\frac{5}{2}^-[532]$ &        & 1726 & 1451 & 275 \\
    $^{160}$Sm & $6^+$   & $\pi^2\frac{5}{2}^-[532]\otimes\frac{7}{2}^-[523]$ &        & 2283 & 2239 & 44  \\
    $^{160}$Sm & $4^+$   & $\pi^2\frac{3}{2}^+[411]\otimes\frac{5}{2}^+[413]$ &        & 2449 & 1772 & 677 \\
    \hline

    $^{162}$Sm & $4^-$   & $\nu^2\frac{1}{2}^-[521]\otimes\frac{7}{2}^+[633]$ &        & 1342 & 1243 & 99  \\
    $^{162}$Sm & $3^+$   & $\nu^2\frac{1}{2}^-[521]\otimes\frac{5}{2}^-[523]$ &        & 1906 & 2015 & 109 \\
    $^{162}$Sm & $6^-$   & $\nu^2\frac{5}{2}^-[512]\otimes\frac{7}{2}^+[633]$ &        & 2201 & 1827 & 374 \\
    $^{162}$Sm & $6^-$   & $\nu^2\frac{5}{2}^-[523]\otimes\frac{7}{2}^+[633]$ &        & 2281 & 2131 & 150 \\
    $^{162}$Sm & $3^-$   & $\nu^2\frac{1}{2}^-[521]\otimes\frac{5}{2}^+[642]$ &        & 2396 & 2331 & 65  \\
    $^{162}$Sm & $3^+$   & $\nu^2\frac{1}{2}^-[521]\otimes\frac{5}{2}^-[512]$ &        & 2496 & 1957 & 539 \\
    $^{162}$Sm & $5^-$   & $\pi^2\frac{5}{2}^+[413]\otimes\frac{5}{2}^-[532]$ &        & 1459 & 1128 & 331 \\
    $^{162}$Sm & $4^-$   & $\pi^2\frac{3}{2}^+[411]\otimes\frac{5}{2}^-[532]$ &        & 1778 & 1444 & 334 \\
    $^{162}$Sm & $6^+$   & $\pi^2\frac{5}{2}^-[532]\otimes\frac{7}{2}^-[523]$ &        & 2211 & 2127 & 84  \\
    $^{162}$Sm & $4^+$   & $\pi^2\frac{3}{2}^+[411]\otimes\frac{5}{2}^+[413]$ &        & 2464 & 1741 & 723 \\
   \hline
\end{longtable}

\subsection{Rotational properties in Nd and Sm isotopes}

\begin{figure*}[h]
\includegraphics[width=0.95\columnwidth]{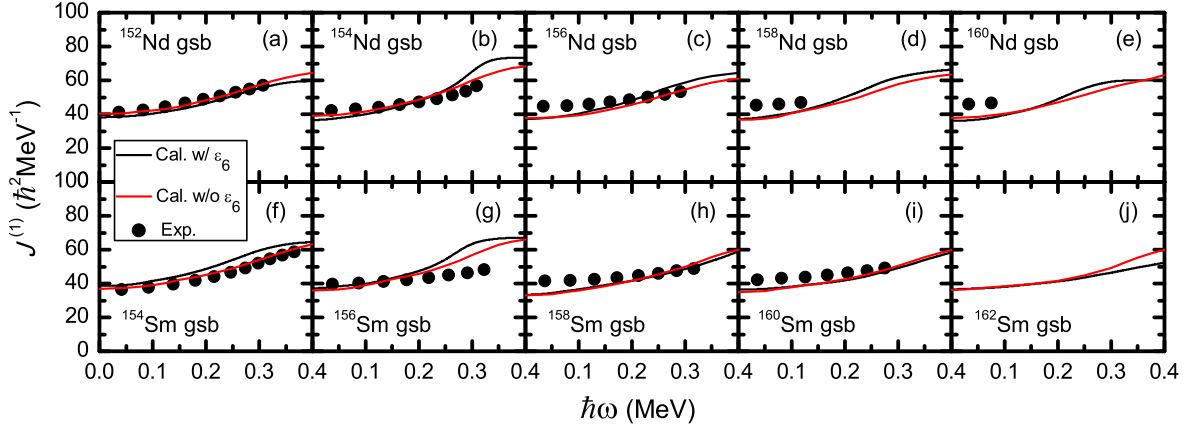}
\caption{\label{fig2:ee}(Color online) The experimental (solid circles) and calculated (solid black lines)
kinematic MOIs $J^{(1)}$ for the GSBs in even-even Nd and Sm isotopes from $N=92$ to $N=100$.
The data are taken from Refs.~\cite{Zhang1998_PRC57-2040, Simpson2009_PRC80-024304, Ideguchi2016_PRC94-064322}.
The calculated results without high-order deformation $\varepsilon_6$ are also shown as red lines.
}
\end{figure*}

Furthermore, the rotational bands observed in Nd and Sm isotopes are analyzed.
Figure~\ref{fig2:ee} shows the experimental (solid circles) and calculated (solid black lines)
kinematic MOIs $J^{(1)}$ for the ground state bands (GSBs) in even-even Nd (upper panel) and Sm (lower panel) isotopes from $N=92$ to $N=100$.
The calculated results without high-order deformation $\varepsilon_6$ are also shown as red lines.
The experimental kinematic MOIs for each band are extracted by
\begin{equation}
\frac{J^{(1)}(I)}{\hbar^2}=\frac{2I+1}{E_{\gamma}(I+1\rightarrow
I-1)}
\end{equation}
separately for each signature sequence within a rotational band
($\alpha = I$ mod 2). The relation between the rotational frequency
$\omega$ and nuclear angular momentum $I$ is
\begin{equation}
\hbar\omega(I)=\frac{E_{\gamma}(I+1\rightarrow
I-1)}{I_{x}(I+1)-I_{x}(I-1)} \ ,
\end{equation}
where $I_{x}(I)=\sqrt{(I+1/2)^{2}-K^{2}}$, $K$ is the projection of
nuclear total angular momentum along the symmetry $z$ axis of an
axially symmetric nuclei.
It can be seen that the MOIs and their variations with the rotational frequency
are well reproduced by the PNC-CSM calculations.
The data show that there is no sharp upbending in all Nd and Sm isotopes,
which is consistent with the PNC-CSM calculations except for $^{154}$Nd and $^{156}$Sm.
It can be seen that obvious upbendings exist in the calculated MOIs of $^{154}$Nd and $^{156}$Sm
when $\varepsilon_6$ deformation is considered.
After the $\varepsilon_6$ deformation being switched off,
the upbendings become less prominent and the results are more consistent with the data.
This indicate that with the rotational frequency increasing,
the $\varepsilon_6$ deformation may become smaller.
It also can be seen that at the low rotational frequency region,
$\varepsilon_6$ deformation has little effect on the MOIs,
while with increasing rotational frequency, it will change the behavior of MOIs.
This is because the $\varepsilon_6$ deformation will change the position of the high-$j$ orbitals,
and then influence the alignment process of these orbitals in high-spin region~\cite{Zhang2013_PRC87-054308, Liu2012_PRC86-011301R}.
Note that the deformation parameter $\varepsilon_6$ is
fixed in the present PNC-CSM calculation,
while it may change with the rotational frequency.
I expect that after considering this effect, the results can be improved further.

\begin{figure*}[h]
\includegraphics[width=0.95\columnwidth]{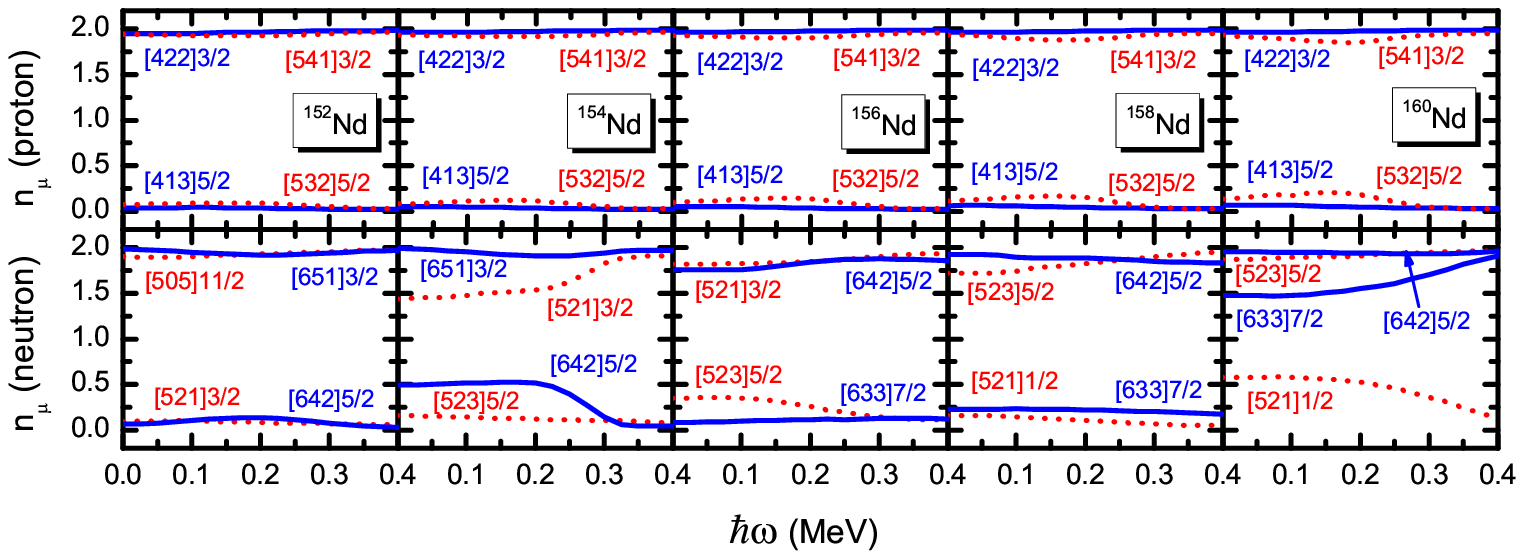}
\caption{\label{fig3:occup}(Color online)
Occupation probability $n_\mu$ of each orbital
$\mu$ (including both $\alpha=\pm1/2$) near the Fermi surface
for the GSBs in Nd isotopes.
The top and bottom rows are for protons and neutrons, respectively.
The positive (negative) parity levels are denoted by blue solid (red dotted) lines.
The Nilsson levels far above the Fermi surface
($n_{\mu}\sim0$) and far below ($n_{\mu}\sim2$) are not shown.
}
\end{figure*}

The experimental MOIs of even-even Nd and Sm isotopes show that the
upbending is weak and becomes less and less obvious with increasing neutron number.
To understand this, the occupation probability $n_\mu$ of each orbital
$\mu$ (including both $\alpha=\pm1/2$) near the Fermi surface
for the GSBs in Nd isotopes is shown in Fig.~\ref{fig3:occup}.
The top and bottom rows are for protons and neutrons, respectively.
The positive (negative) parity levels are denoted by blue solid (red dotted) lines.
The Nilsson levels far above the Fermi surface
($n_{\mu}\sim0$) and far below ($n_{\mu}\sim2$) are not shown.
In the PNC-CSM, the total particle number $N = \sum_{\mu}n_\mu$ is exactly conserved,
whereas the occupation probability $n_\mu$ for each orbital varies
with rotational frequency.
By examining the $\omega$-dependence of the orbitals near the Fermi surface,
one can get some insights on the band crossing.
It can be seen from the upper panel of Fig.~\ref{fig3:occup} that for the proton of $^{152}$Nd,
the orbitals above the Fermi surface are nearly all empty, and
the orbitals below the Fermi surface are nearly all occupied,
and they are nearly unchanged with increasing rotational frequency.
This is due to the large shell gap at $Z=60$, which makes the proton pairing correlations very weak.
While with neutron number increasing, especially for $^{160}$Nd,
$\pi 5/2^-[532]$ and $\pi 3/2^-[541]$ become partly occupied and partly empty, respectively.
This is caused by the decreasing of the $Z=60$ shell gap with increasing neutron number.
With rotational frequency increasing, the occupation of $\pi 5/2^-[532]$ and $\pi 3/2^-[541]$
becomes nearly occupied and nearly empty at $\hbar\omega \approx 0.25$~MeV, respectively.
Therefore, these two proton $h_{11/2}$ high-$j$ orbitals may contribute to the upbending.
It can be seen from the lower panel of Fig.~\ref{fig3:occup} that
for neutrons, only the occupation probabilities of $\nu5/2^+[642]$ in $^{154}$Nd changes drastically
around the upbending frequency. So this neutron $i_{13/2}$ orbital may contribute to the upbending in $^{154}$Nd.
While in other Nd isotopes, the occupation probabilities of all the
orbitals either keep unchange or change gradually with increasing rotational frequency.
Therefore, the contribution to the upbending from neutron in these nuclei may be little.
The present calculations show that the proton $Z=62$
shell gap is smaller than the $Z=60$ shell gap.
Therefore, the proton occupation probabilities of Sm isotopes
must be a little different from those in Nd isotopes. While for neutrons,
the occupation probabilities are very close to each other
when the nuclei with the same neutron number being considered.
Moreover, it can be seen from Fig.~\ref{fig2:ee}
that the behavior of the MOIs for Nd and Sm are quite similar,
so only the occupation probabilities for Nd isotopes are given to illustrate the alignment process.

\begin{figure*}[h]
\includegraphics[width=0.95\columnwidth]{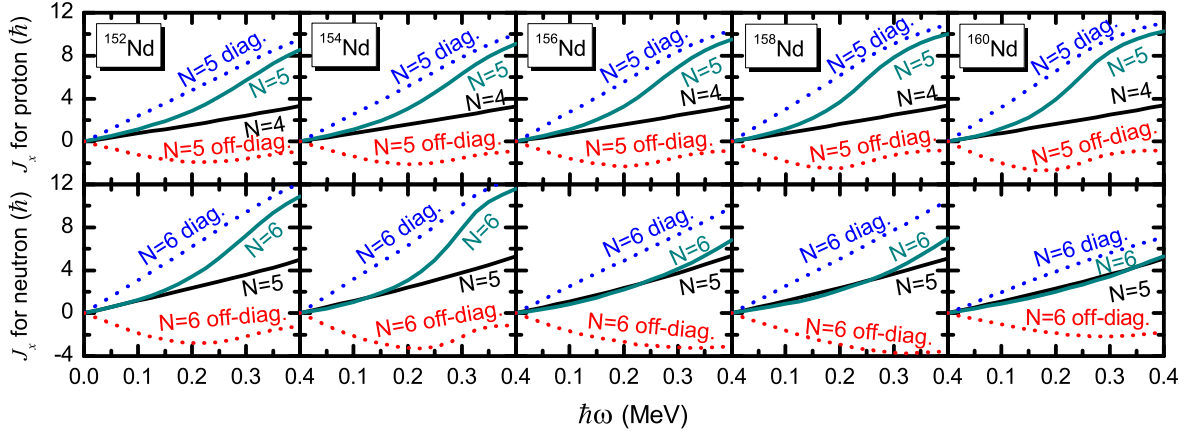}
\caption{\label{fig4:jx}(Color online) Contribution of each proton (upper panel)
and neutron (lower panel) major shell to the angular momentum alignment $ J_x$
for the GSBs in Nd isotones.
The diagonal $\sum_{\mu} j_x(\mu)$ and off-diagonal parts $\sum_{\mu<\nu} j_x(\mu\nu)$
in Eq.~(\ref{eq:jx}) from the proton $N=5$ and neutron $N=6$ shells are shown by dotted lines.
}
\end{figure*}

It is well known that the upbending is caused by the alignment of the high-$j$
intruder orbitals~\cite{Stephens1972_NPA183-257},
which corresponds to the neutron $i_{13/2}$ and proton $h_{11/2}$ orbitals
in rare-earth nuclei.
In order to have a  more clear understanding of the alignment mechanism in these neutron rich nuclei,
the contribution of each proton and neutron major shell to the total angular momentum alignment $ J_x$
for the GSBs in Nd isotopes are shown in Fig.~\ref{fig4:jx}.
It can be seen that for proton, the main contribution to the angular momentum alignment comes
from the $N=5$ major shell ($h_{11/2}$ orbitals).
Moreover, the contribution gradually increases with increasing neutron number.
While for neutron, the contribution from $N=6$ major shell ($i_{13/2}$ orbitals) is prominent only in
$^{152}$Nd and $^{154}$Nd. In $^{156,158,160}$Nd, the contribution from $N=6$ major shell
gets as smaller as  $N=5$ major shell.
This is due to the fact that with neutron number increasing, the high-$j$ but high-$\Omega$
orbital $\nu7/2^+[633]$ gets close to the Fermi surface,
which contributes not very much to the alignment.
Therefore, one can get that different from a typical nucleus,
in which the upbending is caused by whether the neutron or the proton alignment,
both neutron and proton alignments contribute
to the upbending in these neutron rich Nd and Sm isotopes.
In the lighter Nd and Sm isotopes, the alignment
is due to both neutron $i_{13/2}$ and proton $h_{11/2}$ orbitals.
Meanwhile, the proton $h_{11/2}$ orbitals play a more and more important role in
the alignment process with neutron number increasing.
The competition between the alignment of proton and neutron high-$j$ orbitals
makes the upbending in these Nd and Sm isotopes
very weak and less obvious with increasing neutron number.

\begin{figure*}[h]
\includegraphics[width=0.95\columnwidth]{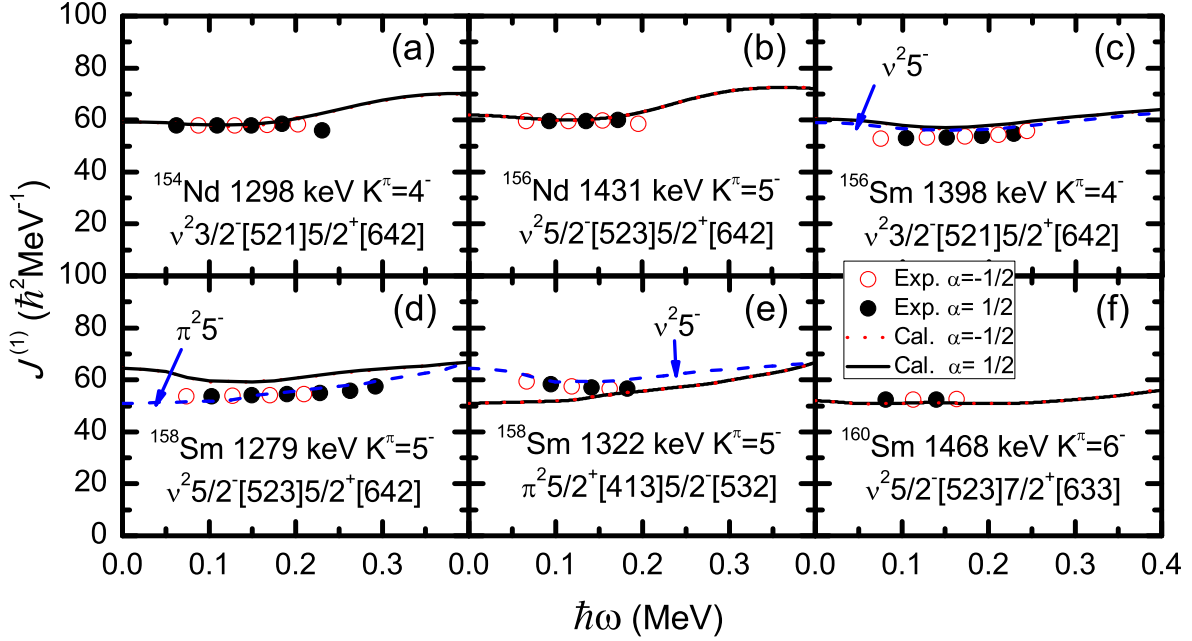}
\caption{\label{fig5:2qp}(Color online)
The experimental and calculated MOIs of 2-qp bands in Nd and Sm isotopes.
The experimental MOIs are denoted by full black cicles
(signature $\alpha=0$) and open circles (signature $\alpha=1$), respectively.
The calculated MOIs by the PNC-CSM are denoted by black solid lines
(signature $\alpha=0$) and red dotted lines (signature $\alpha=1$), respectively.
The data are taken from Refs.~\cite{Simpson2009_PRC80-024304, Wang2014_PRC90-067306, Patel2016_PLB753-182}.
$\nu^2 5^-$ and $\pi^2 5^-$ denote $\nu^2 5/2^-[523]\otimes 5/2^+[642]$ and
$\pi^2 5/2^+[413]\otimes 5/2^-[532]$, respectively.}
\end{figure*}

Figure~\ref{fig5:2qp} shows the experimental and calculated MOIs of 2-qp bands in Nd and Sm isotopes.
The experimental MOIs are denoted by full black cicles
(signature $\alpha=0$) and open circles (signature $\alpha=1$), respectively.
The calculated MOIs by the PNC-CSM are denoted by black solid lines
(signature $\alpha=0$) and red dotted lines (signature $\alpha=1$), respectively.
The data are taken from Refs.~\cite{Simpson2009_PRC80-024304, Wang2014_PRC90-067306, Patel2016_PLB753-182}.
It can be seen that the data can be reproduced very well by the PNC-CSM calculations,
except two $K^\pi = 5^-$ bands in $^{158}$Sm.
The agreement between the calculation and the data also supports the configuration
assignments for these 2-qp states.
Note that in Refs.~\cite{Simpson2009_PRC80-024304, Wang2014_PRC90-067306}, the 1279~keV level is assigned as
$\nu^2\ 5/2^-[523]\otimes5/2^+[642]$ and the 1322~keV level is assigned as
$\pi^2 5/2^+[413]\otimes 5/2^-[532]$, respectively.
However, the present PNC-CSM calculations show that,
if the configuration assignments for these
two bands are changed, the MOIs can be reproduced quite well.
In addition, as I mentioned before,
the 1398~keV state in $^{156}$Sm was previously assigned
as $K^\pi=5^-$ in Ref.~\cite{Simpson2009_PRC80-024304}.
It can be seen that the calculated MOIs for this band with
$K^\pi = 4^-$ and $5^-$ configurations are similar.
Therefore, it is difficult to distinguish these two configuration assignments by their MOIs.
Due to the fact that the calculated excitation energy of $K^\pi = 4^-$ is very close to the data,
this state is attentively assigned as $K^\pi = 4^-$.
More detailed experimental information is needed to give a solid configuration assignment for this state.

\begin{figure*}[h]
\includegraphics[width=0.95\columnwidth]{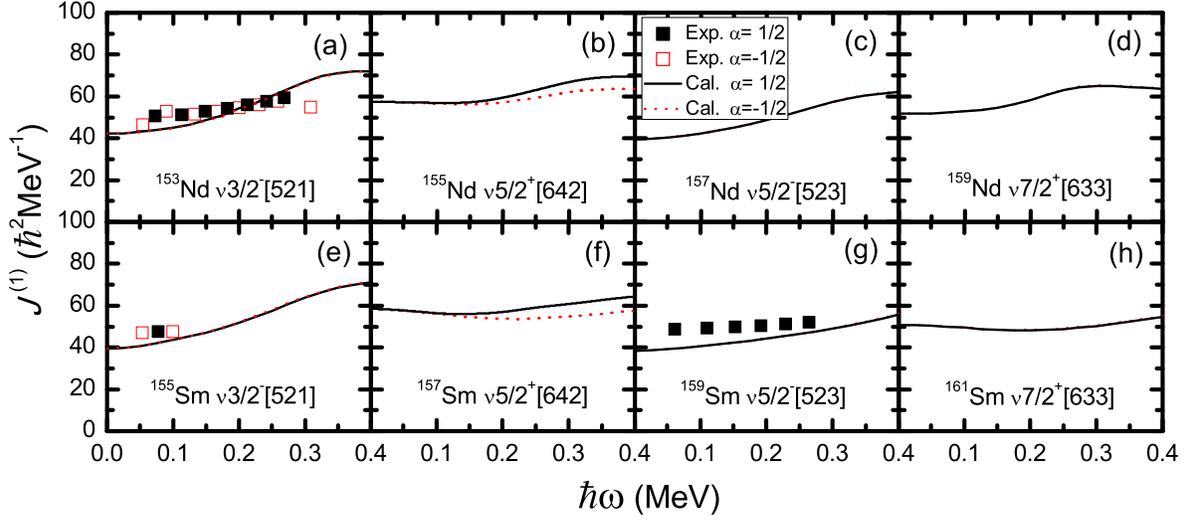}
\caption{\label{fig6:oa}(Color online) The experimental and calculated kinematic MOIs $J^{(1)}$
of the GSBs for odd-$A$ Nd and Sm isotopes.
The data are taken from Refs.~\cite{Hwang1997_IJMPE6-331, Reich2005_NDS104-1, Urban2009_PRC80-037301}.
The experimental MOIs are denoted by full square
(signature $\alpha=+1/2$) and open square (signature $\alpha=-1/2$), respectively.
The calculated MOIs by the PNC-CSM are denoted by solid lines
(signature $\alpha=+1/2$) and dotted lines (signature $\alpha=-1/2$), respectively.
}
\end{figure*}

\begin{figure*}[h]
\includegraphics[width=0.95\columnwidth]{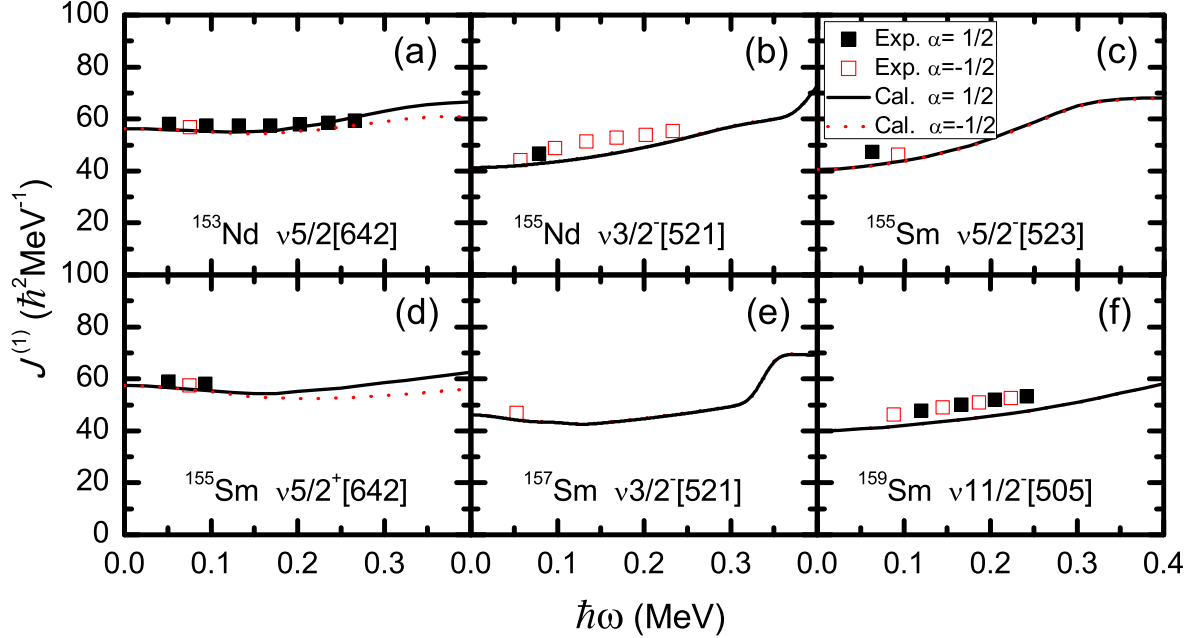}
\caption{\label{fig7:oa}(Color online) Similar as Fig.~\ref{fig6:oa},
but for the excited 1-qp bands in odd-$A$ Nd and Sm isotopes.
The data are taken from Refs.~\cite{Reich2005_NDS104-1, Hwang2008_PRC78-014309,
Urban2009_PRC80-037301, Hwang2010_PRC82-034308}.
}
\end{figure*}

Except even-even Nd and Sm isotopes, some 1-qp rotational bands,
including both the ground and the excited state bands, have been identified in the
odd-$A$ nuclei, which can provide more detailed information on the single-particle structure
for these neutron rich nuclei.
Figure~\ref{fig6:oa} shows the experimental and calculated kinematic MOIs $J^{(1)}$
of the GSBs for odd-$A$ Nd and Sm isotopes.
The available data are taken from Refs.~\cite{Hwang1997_IJMPE6-331, Reich2005_NDS104-1, Urban2009_PRC80-037301}.
The experimental MOIs are denoted by full square
(signature $\alpha=+1/2$) and open square (signature $\alpha=-1/2$), respectively.
The calculated MOIs by the PNC method are denoted by solid lines
(signature $\alpha=+1/2$) and dotted lines (signature $\alpha=-1/2$), respectively.
The experimental data show that the ground states of $N=153$ ($^{153}$Nd and $^{155}$Sm)
and $N=157$ ($^{157}$Nd and $^{159}$Sm) isotones are $\nu3/2^-[521]$ and $\nu5/2^-[523]$, respectively.
The present calculations are also consistent with the data.
Therefore, from the cranked Nilsson levels in Fig.~\ref{fig1:Nil} we can get
that the ground state of $N=155$ isotones should be $\nu^-5/2[642]$.
However, in Ref.~\cite{Hwang2008_PRC78-014309} the data show that the ground
state of $^{155}$Nd is $\nu3/2^-[521]$, which is different from the PNC-CSM calculation.
From a systematic point of view, I put the $\nu^-5/2[642]$ as the ground state of $N=155$ isotones.
Similar as Fig.~\ref{fig6:oa}, the MOIs for the 1-qp excited bands
are shown in Fig.~\ref{fig7:oa}.
The data are taken from Refs.~\cite{Reich2005_NDS104-1, Hwang2008_PRC78-014309,
Urban2009_PRC80-037301, Hwang2010_PRC82-034308}.
It can be seen from Figs.~\ref{fig6:oa} and~\ref{fig7:oa} that the MOIs
of these rotational bands in odd-$A$ Nd and Sm isotopes can be well reproduced by the PNC-CSM,
which in turn confirms the configuration assignments for these 1-qp states.

\section{\label{Sec:Summary}Summary}

In summary, the rotational properties of the neutron rich Nd and Sm isotopes
with mass number $A\approx150$ are investigated using the cranked shell model
with pairing correlations treated by a particle-number conserving method,
in which the Pauli blocking effects are taken into account exactly.
The excitation energies of several experimentally observed 2-qp
$K$ isomers are reproduced quite well by the PNC-CSM calculation.
Furthermore, all 2-qp states in even-even Nd and Sm isotopes with excitation energies lower
than 2.5~MeV are systematically calculated,
and possible 4-qp $K$ isomers with the lowest 2-quasi-proton and
2-quasi-neutron configurations are predicted.
Moreover, the experimentally observed rotational frequency variations of MOIs
for the even-even and odd-$A$ nuclei
are reproduced very well by the PNC-CSM calculations.
The effects of high-order deformation $\varepsilon_6$ on the 2-qp
excitation energies and MOIs of the GSBs are analyzed.
By analyzing the occupation probability $n_\mu$ of each cranked
Nilsson orbitals near the Fermi surface and the contribution of
each major shell to the angular momentum alignments,
the alignment mechanism in these nuclei is understood clearly.

\section{Acknowledgement}

Helpful discussions with Shan-Gui Zhou are gratefully acknowledged.
This work was partly supported by the National Natural Science
Foundation of China (Grants No. 11505058, 11775112, 11775026),
and the Fundamental Research Funds for the Central Universities (2018MS058).


%

\end{document}